\documentclass{elsart}

 \usepackage{epsfig}

\usepackage{amssymb}

\begin{document}

\begin{frontmatter}



\title{BCS Model in Tsallis' Statistical Framework}


\author{Lizardo H. C. M. Nunes }\ead{lizardo@if.uff.br}\and\author{E. V. L. de Mello}
\address{ Departamento de F\'{\i}sica, Universidade Federal Fluminense,
          Av. Litor\^anea s/n, 
	  Boa Viagem
	  Niter\'oi, Rio de Janeiro,
          24210-340,
          Brazil }

\begin{abstract}
We show that there is an effect of nonextensivity 
acting upon the BCS model for superconductors in the ground state that motivates its study in
the Tsallis' statistical framework. We show that the weak-coupling limit superconductors
are well described by $ q \sim 1 $, where $ q $ is a real parameter which characterizes the degree
of nonextensivity of the Tsallis' entropy.
Nevertheless, small deviations with respect to $ q = 1 $ 
provide better agreement when compared with experimental results.
To illustrate this point, making use of an approximated Fermi function,
we show that measurements of the specific heat, ultrasonic attenuation
and tunneling experiments for tin (Sn) are better described with $ q = 0.99 $.
\end{abstract}

\end{frontmatter}

\section{Introduction}
\label{Int}
About 10 years ago, Tsallis\cite{Tsallis} proposed the following entropic 
form:
\begin{equation} \label{3.1.1}
S_q = -k_B \frac{ 1 - Tr {\bf \rho }^q }{ 1 - q } \, ,
\end{equation}
where $ k_B $ is a positive constant, $ { \bf \rho } $ is the matrix density 
of the system and $ q $ is a real parameter.
The thermodynamics constructed from (\ref{3.1.1})
has been applied to several physical systems in the last years.
Indeed, nonextensive statistics was used to investigate physical systems 
which presented nonextensive features. Amongst them, there are stellar 
polytrops\cite{PlastinoPlastino1990}, solar neutrino problem\cite{Kaniadakisetal},
galaxy clusters\cite{Lavagnoetal}, etc.
It should be remarked that the formalism established above seems to be 
appropriate particularly to systems interacting via long-range interactions.

Furthermore, formal results were achieved and techniques were developed. Indeed, from the 
properties of the Tsallis' entropy, Plastino and Tsallis\cite{PlastinoeTsallis}
generalized the Boguljubov inequalities and justified the variational 
method; also, this nonextensive thermodynamics preserves the Legendre 
Transform formalism\cite{PlastinoePlastino2,CuradoeTsallis}.
It's also possible to establish a relation between the partition 
function with $ q = 1 $, as usually calculated by the standard statistics,
and the generalized $ q \not= 1 $ nonextensive partition function\cite{Prato}
(for this purpose one use the Hilhorst integral ). This result was crucial to 
the generalization of the Green's functions 
method\cite{MendesLenzieRajagopal,Lenzietal}, which 
is a useful tool in the discussion of a many-body system for
any possible real value for $ q $, where $ q $ measures the
nonextensivity of the system.

In this article study the
BCS (Bardeen, Cooper and Schrieffer) model in Tsallis' statistical framework.
Indeed, the BCS theory for superconductivity describes very well, at least qualitatively,
a vast amount of experimental results
for a wide range of materials; these materials, henceforth, will be called usual superconductors, since
they are well explained by the BCS theory.
Assuming that the exchange of phonons between
electrons is responsible for the superconducting phenomena, we shall also assume that the criterium
for the appearance of superconductivity is that the Coulomb repulsion is surmounted by the
effective attraction caused by the phononic interaction.
Instead of calculating the interactions involved by ``first principles''
(divergences in the perturbative chain postponed for years the development of the theory)
BCS simply considered a mean-field approach with excellent results.
Such improved theory will not be considered in this article.
In fact, our aim is to develop a $ q $-dependent formalism which may be applied to materials
or a class of materials which systematically deviates from the usual theory. As a first approach,
we will concern ourselves to the weak-coupled limit of the BCS theory, but the strong coupling
limit and the generalization of the theory to other potentials with the correlations included,
for instance, is already under study and will be subject of further publication.

In next section we show that there is an effect of nonextensivity acting upon the BCS model
at $ T = 0 $, then we generalize the well known gap equation and use it to calculate
quantities that will be compared with the experimental results for the usual superconductors.

\section{Motivation}
\label{Mot}

Consider the well known ``reduced'' Hamiltonian proposed by BCS and used to 
study the appearance of superconductivity in simple metals \cite{BCS}, 
\begin{equation}
    H_{\mbox{red}} = \sum_{k} \epsilon_{k} \left( n_{k} + n_{-k} 
    \right) + \sum_{k k'} V_{kk'} b^{\dagger}_{k'} b_{k} \, ,
    \label{Hred}
\end{equation}
where the operator $ b^{\dagger}_{k } = c^{\dagger}_{k }c^{\dagger}_{-k } $ 
creates a Cooper pair in the singlet state and the model describes 
interacting electrons forming singlet Cooper pairs.
This assumption may be justified
in terms of an instability of the Fermi sea being caused by 
binding of zero-momentum pairs. The only relevant electrons to the 
superconducting state are those 
in a small neighborhood of the Fermi level within a layer twice the 
Debye frequency thick (If one consider electron-phonon interaction as 
the sole responsible for the superconductivity phenomena.). Notice 
that we 
have used the convention that an explicitly negative $ k $ has down 
spin and vice versa.

The transformed Hamiltonian, in 
a semi-classical view, has its unperturbed Fermi 
sea characterized by ``spins'' aligned only in 
the $\hat{z}$ direction; once the 
phononic attraction between electrons is present, a non-null component of $ \vec{ 
S_{k} } $ perpendicular to $\hat{z}$ shows up and this configuration 
is determined in terms of the orientation $ S_{ x k } / S_{ z  k} = 
\tan\theta_{k} $. Indeed, following 
Bogoliubov's notation\cite{Bogoljubov}, 
$\sin\theta_{k}=u_{k}v_{k} $ and this quantity is related to the 
superconducting gap energy, $\Delta_{k} $, this gap may be seen as the order
parameter of the phase transition superconductor-normal 
($ \Delta_{k} \neq 0 $ represents superconducting solutions.).

Following the prescription briefly introduced above,
the system may be analyzed not at an arbitrary 
temperature $ T $ below the transition temperature  $ T_{c} $, but 
particularly at $ T = 0 $, where all relevant electrons are in the 
condensate state, henceforth, paired. When the temperatures increases, 
the possible excitations of the system are the quasi-particles and 
the excited pairs. As pointed out by Schrieffer\cite{Schrieffer}, 
at the zero-temperature the operator $ n_{k} +
n_{-k} $ may be replaced by $ 2 b^{\dagger}_{k} b_{k} $, twice the 
occupation number of pairs. Hence, we have to work in the grand-canonical
ensemble adding to (\ref{Hred}) the operator $- \mu N $, where $ N $ is 
$ \sum_{k} n_{k} + n_{-k} $ and the chemical potential, $ \mu $, may 
be regarded as essentially the Fermi energy\cite{Tinkham}. Therefore,
we get a modeled system of interacting electrons with net energy $ 
\epsilon_{k} - \mu $, what is consistent with the picture of only 
relevant electrons for superconductivity possessing energies in the vicinity of the Fermi level.

In order to study the attractive potential, we perform a
Fourier transform of (\ref{Hred})
assuming an isotropic superconductor. The 1/2 ``spin'' operators are
defined by
\begin{eqnarray}
b^{\dagger}_{i} & = & S_{ i x } + i S_{iy}  =  S^{+}_{i}  \, ; 
    \label{S+}  \\
b_{j} & = & S_{ j x } - i S_{ j y }  =  S^{-}_{j} \, .
    \label{S-} 
\end{eqnarray}
After writing the ``reduced'' Hamiltonian at $ T = 0 $ in terms of these 
pseudo spins(\ref{S+}, \ref{S-}),it is possible to show that the
model is isomorphous to the $ X Y $ model:
\begin{equation} \label{XY}
{\bf H}_{red} = \sum_{ij} ( t_{ij} + V_{ij}  ) b^{\dagger}_{i} b_{j} \, ,
\end{equation}
where
\begin{equation} \label{tij}
t_{ij} = \frac{2}{\Omega} \sum_{k} \left( \epsilon_{k} - \mu
\right) e^{-i {\bf k} ( {\bf 
r_{i}}- {\bf 
r_{j} } ) }   \, 
\end{equation}
and
\begin{equation} \label{Vij}
V_{ij} = \frac{2}{\Omega} \sum_{kk'} V_{kk'}  e^{-i ( {\bf k r_{i}}- {\bf 
k'
r_{j} } ) }   \, ,
\end{equation}
with $\Omega $ being a normalizing constant.

As pointed out by Tsallis and 
Anteneodo \cite{AnteneodoeTsallis}, this 
classical Hamiltonian may be studied with
$ V_{ij} \propto r_{ij}^{-\alpha} $
as the coupling between sites, where $ r_{ij} = r_{i }- r_{j} $ 
and $ \alpha $ is the range of the interaction essentially. A 
typical quantity that should be calculated at $ T = 0 $ is 
$\int_{1}^{\infty} dr \, r^{ d -1 } r^{ - \alpha } $,
which is related to the internal energy per particle. 
If $ 0 \le \alpha \le d $, It's a trivial task to show that the 
internal energy per particle diverges in the 
thermodynamic limit, what was called weak 
violation of Boltzmann-Gibbs (BG) statistics (in contrast to 
``strong'' $T \neq 0 $ violation). This result also reveals the role played by the
range and the dimension of the system in the scenario of the
nonextensivity.

Following the path established above and assuming $ d = 
3 $, we will show that there is a nonextensive effect 
acting upon the BCS model at $ T = 0 $. For this purpose we will use
the well known BCS mean-field interaction
\begin{equation} \label{BCSint}
V_{kk'} =  \left\{
    \begin{array}{l}
     -V \; , \mbox{ we }|\epsilon_k - \mu|, \,|\epsilon_{k'} - 
\mu| < \hbar w_D \\
      0 \; , \mbox{ otherwise }
    \end{array}  \right. \, ,
\end{equation}
where $\hbar w_D = k_B \theta_D $ and $\theta_D$ 
is the Debye temperature. The spatial dependence of this interaction 
will be found approximating the sums over $ k $ and $ k' $ in 
(\ref{Vij}) by volume integrals. The region of integrability, of course, 
will be constrained to values of $ {\bf k } $ related to the relevant
electrons only: 
$ \alpha_D \equiv \sqrt{ \frac{2 m_e w_D}{\hbar} } > | | {\bf k } | - k_F | $,
where $ m_e $ is the electron mass, what defines a neighborhood around $ 
k_{F } $. Hence, we finally get
\begin{equation}
   V_{ij}  = \frac{ (8 \pi)^2 }{\Omega} \, v( r_{\beta} )  \, , 
    \label{VijBCS}
\end{equation}
with
\begin{eqnarray} 
  v( r_{\beta} ) & = &  \frac{ 1 }{ r^{2}_{ \beta } } \, [ \, \frac{ \cos(k_F 
                           r_{\beta}) \sin(\alpha_D r_{\beta}) }{r_{\beta} } - \nonumber \\
                 &   & \alpha_D \cos(\alpha_D r_{\beta} )
\cos(k_F r_{\beta}) + k_F \sin(k_F r_{\beta})\sin(\alpha_D r_{\beta}) \,]   
    \label{vbeta}
\end{eqnarray} 
and $ \beta = i,j $ indistinctly (Notice $ r_{\beta}= |{\bf r_{\beta} }| $).

There's a one-to-one correspondence between the $ i$'s lattice sites 
and the pairs at $ T = 0 $, so it's reasonable to replace sums 
over $ i $ and $ j $ by integrals. Again, we get a 
product of identical integrals and the potential energy is 
proportional to $ [\, \int \, d^{3} r \, v_{ \beta } \,]^{2} $.
In agreement of what is expected for a extensive system, 
the first part of the above integral converges to a constant factor
in the thermodynamic limit  ( since 
$ k_F, a_D > 0 $ and  $\lim_{N \to \infty} \, \int d^{3}r \, 
\left( \cos(k_F r) \, \sin(\alpha_D r) / r^3 \right) = \pi /2 $\, ). However, the 
second and third contribution give raise to oscillatory fluctuations in the 
potential energy according to the number of pairs in the system:  
\begin{eqnarray} \label{2.3.13}
\lefteqn{ \int \, d^3r \, \frac{ 1 }{ r^{2} }  \, 
             \left[\, k_F  \sin(k_F r) \sin(\alpha_D r)  - 
	     \alpha_D  \cos(\alpha_D r ) \cos(k_F r)   \, \right] =} \nonumber \\
        & &  4 \pi \left\{ \, \sin[ ( k_F + \alpha_D )N ] -
	                      \sin[ ( k_F - \alpha_D )N ] \, \right\} \, . 
\end{eqnarray} 
These fluctuations are bounded but still present even for a very 
large number of pairs and they don't decrease as you add electrons to 
the system. Further, the energy cannot be expressed as the energy per 
particle times the number of particles itself, as commonly done for 
extensive systems. 

Therefore, we conclude that 
there is a formal {\em nonextensive effect acting upon the BCS model at $ T = 0 $}
as mentioned above. The nonextensive character may be regarded as a 
motivation for the analysis of the BCS theory in Tsallis' statistical 
framework.

\section{The Method}
\label{Method}

In this section we will generalize the BCS gap
equation in the Tsallis' statistical framework
making use of the well known Gorkov's anomalous
functions. Also, it's that the lifetime
of the quasi-particles is infinity,
since the interaction is time-independent.

Originally, BCS proposed a wave function
for the ground state,
$ | \psi_{0} \rangle $, describing $ N $ paired electrons.
This wave function was given in terms of a variational
parameter that should be found minimizing the Helmholtz free energy.
Regarding BCS, the superconducting gap energy, $ \Delta $,
may be considered the order parameter of the phase transition. 
Notice that even before the full development of the theory this gap energy was 
observed in several experiments such as specific heat,
ultrasonic attenuation rate, tunneling measurements, etc.
The self-consistent equation for the superconducting gap was given by BCS\cite{BCS} in a
Hartree-Fock approximation and may be written as
$ \Delta_k = - \sum_{kk' } V_{kk'} \langle b^{\dagger}_{k'} \rangle \,$, where
$ \langle  b^{\dagger}_{k} \rangle $ is the expectation value in the 
BG statistics for the temperature dependent
creation operator of a Cooper pair. A generalized version of this quantity,
which includes Tsallis statistics and the $ q $-dependent expectation value,
shall simply be
\begin{equation}
    \Delta_k = - \sum_{kk' } V_{kk'} \langle b^{\dagger}_{k'} \rangle_q \, .
    \label{gap}
\end{equation}
Through a canonical transformation the system may be simply described
by quasi-particles with a given dispersion relation\cite{Bogoljubov}
In order to calculate the $ q $-expectation value found in (\ref{gap})
we define , for $ q = 1 $ and except for a multiplicative
factor, the propagator
$G_{\sigma}({\bf k}, \tau) = \langle \, T [c_{k \sigma }(\tau)
\, c^{\dagger}_{k
\sigma}(0) ] \, \rangle $, where $ T $ is the temporal ordering
operator and
$ \hbar =1 $ from now on. The propagator can be rewritten as
$ G_{\sigma}({\bf k}, \tau) = \theta(\tau) \langle \, c_{k \sigma }(\tau) \,
c^{\dagger}_{k \sigma}(0) \rangle -
\, \theta(-\tau) \langle \, c^{\dagger}_{k \sigma}(0)  c_{k \sigma }(\tau) \, \rangle $
in a fermionic system.
In order to calculate $ G $ in interacting systems we may use two
methods mainly: Feynman diagrams and the construction of the
equation of moment. We attempt to generalize the
latter by closely following the development
discussed in Ref\cite{BogoljuboveBogoljubovJr}.

Consider the advanced($ a $) and retarded($ r $) double-time Green's functions
\cite{Mattuck,MendesLenzieRajagopal}:
\begin{eqnarray}
G^{(r,q)}_{AB} = \theta(t - t') \langle \left[ \, {\bf A}(t) , \, {\bf
B}(t')
\, \right]_{\eta } \rangle_q  \label{A.1.5a}\\
G^{(a,q)}_{AB} = - \theta(t - t') \langle \left[ \, {\bf A}(t) , \, {\bf
B}(t')
\, \right]_{\eta } \rangle_q \label{A.1.5b}
\end {eqnarray}
where $\theta(\tau)$ is the  characteristic function times an
arbitrary multiplying factor.
By  definition, $\langle \left[\, {\bf A},\, {\bf B} \,
\right]_{\eta}
\rangle_q  = \langle \, {\bf A} {\bf B} + \eta {\bf B}{\bf A} \,
\rangle_q$, with $\eta= -1$; if $ { \bf A } $ and $ {\bf B }$ are
bosonic operators, and $ \eta= 1 $ if fermionic.
The Green's functions admits a Fourier series expansion
with Fourier transform
\begin{equation} \label{A.1.9}
\ll A, B \gg^{(j,q)}_w = \frac{1}{2 \pi } \int_{-\infty}^{\infty} \, d\tau
\,
G^{(j,q)}_{AB}(t,t') \, e^{iw\tau} \,  ,
\end{equation}
where $ j = r, \, a $ and $ \tau = t - t' $.
Assuming that
$ \left[ \, {\bf A(t)}, \, {\bf B(t')} \, \right]_{\eta} $
is a non null operator, the Green's function can be represented by
means of the Hilhorst integration\cite{MendesLenzieRajagopal,Prato},
which relates the $ q $-expectation value of observables to its
standard $ q = 1 $ analogous average in the BG
statistics. The correlation function is given then by:
\begin{eqnarray} \label{A.1.11}
\lefteqn{ \langle {\bf B}(t') {A}(t) \rangle_q = } \nonumber \\
& & \frac{1}{{\mathcal{Z}}_q } \,
\int_{\mathcal{C}} \, d\xi \, K_q(\xi) {\mathcal{Z}}_1\left(\, (q - 1 )
\xi \bar{\beta}
\,\right) \times \nonumber \\
& &  \int_{-\infty}^{\infty} \, dw \, \frac{ \ll A, B \gg_{w + i0^+}^{(1)}
- \ll A, B \gg_{w - i0^+}^{(1)} }{ e^{ (q-1) \xi \bar{\beta} w } + \eta }
\, ,
\end {eqnarray}
where the kernel $ K_q(\xi) $ is a $ q $-dependent function
presented in the integral representation of the Green's
function, according Lenzi's notation\cite{Lenzietal}.

We shall calculate the equation of motion of each propagator of interest
regarding the phenomena of superconductivity and than calculate the
gap energy, in order to compare it with experimental measurements.

In a time-independent Hamiltonian (such as the case) the
corresponding Green's functions depend only upon $ t - t' = \tau $.
The equation of motion for
$ \ll c_{k \uparrow },c^{\dagger}_{ k \uparrow } \gg^{(1)}_w $,
is:
\begin{equation} \label{A.1.16}
(w - \gamma_k) \ll c_{ k \uparrow },c^{\dagger}_{k \uparrow }
\gg^{(1)}_w = \frac{i}{2\pi} + \Delta
\ll c^{\dagger}_{-k \downarrow }, c^{\dagger}_{k \uparrow}\gg^{(1)}_{w} \, ,
\end{equation}
where $ \gamma_{k} = \epsilon_{k} - \mu $.
For
$\ll c^{\dagger}_{-k \downarrow } c^{\dagger}_{k \uparrow}, \gg^{(1)}_{w} $,
the procedure is identical and
\begin{equation} \label{A.1.15}
( w + \gamma_k) \ll c^{\dagger}_{-k \downarrow},c^{\dagger}_{k \uparrow }
\gg^{(1)}_{w} = \Delta \ll c^{\dagger}_{k \uparrow } , c_{-k \uparrow }
\gg^{(1)}_w \, .
\end{equation}
This is the Gorkov's anomalous function and it's strictly related to the
superconducting phenomena, i. e., the appearance of Cooper
pairs. Notice that the results so far obtained are known and are also restricted
to the standard statistics.

Finally, the propagator
$\ll c^{\dagger}_{-k \downarrow }, c^{\dagger}_{k \uparrow} \gg^{(1)}_{w}$
is
\begin{eqnarray} \label{A.1.18}
\lefteqn{ \ll c^{\dagger}_{k \uparrow},c^{\dagger}_{-k \downarrow }
\gg^{(1)}_{w} =
\left( \frac{i}{2\pi} \right) \frac{\Delta}{w^2 - \gamma_k^2 - \Delta^2 }
} \nonumber \\
& & \hspace{2.2cm} = \left( \frac{i}{2\pi } \right)
\frac{\Delta}{2 E_k} \left[ \, \frac{1}{w - E_k} -
\frac{1}{w + E_k } \, \right] \, ,
\end {eqnarray}
where $ E_k = \sqrt{ (\epsilon_{k } - \mu )^2 + \Delta_k^2 } $
is the energy of the  quasi-particle. When $ \Delta = 0 $,
we get $ E_{k} = \epsilon_{k} - \mu $, which is the electron energy
with respect to the chemical potential $ \mu $ and, indeed,
superconducting solutions are given by $ \Delta \ne 0 $.
Furthermore, from $ E_{k} $, one can easily realize that
the minimum amount of energy to create an excitation in the
superconducting state is $ \Delta $.

Combining (\ref{A.1.11}) and making use of the
$\left( {w - w' \pm i0^+}
\right)^{-1} =
P\left( {w- w'}\right)^{-1} \mp i\pi
\delta\left( w - w' \right)$, one have
\begin{eqnarray} \label{A.1.19}
\lefteqn{ \langle c^{\dagger}_{-k \downarrow }, c^{\dagger}_{k \uparrow}
\rangle_q =} \nonumber \\
& &  -\frac{1}{{\mathcal{Z}}_q } \int_{\mathcal{C}} \, d\xi \, K_q(\xi ) \,
{\mathcal{Z}}_1(\, (q-1) \xi \bar{\beta} \,) \: \frac{\Delta}{2 E_k}
\int_{-\infty}^{\infty} \, dw \, \frac{\delta(w -E_k ) - \delta(w + E_k)}
{e^{(q-1) \xi \bar{\beta} w } + 1} = \nonumber \\
& & -\frac{\Delta}{2 E_k {\mathcal{Z}}_q } \int_{\mathcal{C}} \, d\xi \,
K_q(\xi) \, {\mathcal{Z}}_1(\, (q-1) \xi \bar{\beta} \,) \, \left[ 1 -
2f_1\left( (q-1) \xi \bar{\beta} E_k \right) \right] \, .
\end {eqnarray}
One can easily realize that the whole $ q $-dependence of the quantity
above is contained in the generalized version of the Fermi
distribution function, $ f_{q} $, since
$ {\mathcal{Z}}_{q}^{-1} \,
\int_{\mathcal{C}} \, d\xi \, K_q(\xi)
{\mathcal{Z}}_1 \left(\, (q - 1 ) \xi \bar{\beta} \,\right)  = 1 $
and $f_{q} $ can be also written in terms of its own integral
representation.

Finally, the self-consistent equation for the superconducting gap
equation is:
\begin{equation} \label{A.1.19.1}
\Delta = - \, \sum_{k} V_{kk'} \, \langle b^{\dagger}_k \rangle_q =
  \, \sum_{k} \, V_{kk'} \frac{\Delta}{2 E_k } \left[ 1 - 2f_q({\beta} E_k)
\right] \, ,
\end {equation}
which is identical to that found  by simply diagonalize the reduced
Hamiltonian by a canonical variable transformation proposed by
Bogoljubov\cite{Bogoljubov} but with $ f_q $ in the place of the usual
Fermi function $ f $.

In the case usual superconductors it is reasonable
to assume that the $ q $ involved in numerical calculations
will not deviate appreciably from the unity, since most of their properties have
fairly agreed with the BCS results in the standard statistics.
Hence we may use in numerical calculations an approximated
Fermi function\cite{BuyukkilicDemirhaneGulec},
\begin{equation} \label{fq}
f_q =
\frac{1}{ \left( e_q^{-\beta \gamma_k } \right)^{-q} + 1} \, ,
\end{equation}
where $  e^{x}_q \equiv [ 1 + (1 - q )x ]^{ \frac{1}{1 - q} }  $
is the generalized $ q $-exponential function. The distribution function above
avoids the self consistency found in (\ref{A.1.11}) and
might be considered a useful tool for systems near the extensivity\cite{TirnaklieTorres}.

Following BCS, we replace the sum over $ k $ by an integral over
$ \gamma $ and the density of states shall
be simply written as $ N(\gamma) \approx N( \epsilon_{F} ) $ or $
N(0) $.
Hence, the generalized self-consistency condition is
\begin{equation}
  \frac{1}{N(\epsilon_F)V} = \int_0^{\hbar w_D} \, d\gamma \, \frac{1 - 2
f_q(\beta E)
}{E} \,
    \label{selfcondition}
\end{equation}
and one can estimate the gap energy in the weak-coupling limit
assuming $ T_{c } \ll \theta_{D} $:
\begin{equation} \label{ratio}
\frac{\Delta(0)}{k_B T_c} = 2 e^{- N_q} \, ,
\end {equation}
where
\begin{equation} \label{Nq}
N_q = 2\, \int_{0}^{\infty} dx \, \frac{d}{dx}f_q(x) \, lnx \, ,
\end {equation}
since \cite{BCS}$\Delta(0) = 2 \hbar w_D \, e^{-1 / N(\epsilon_F) V }$.
In the limit $ q \rightarrow 1 $ we get
$ \Delta(0) / k_{B } T_{ c } = 1.76 $, as usual.
If one calculate self-consistently the gap equation taking into account the
experimental data for $ T_{c}  $ and $ \theta_{D} $, one get that the
coupling
$ N(\epsilon_{F} ) V $ is much less than unity. Superconductors with
larger ratio $ T_{c} / \theta_{D} $ tend to have greater disagreement
with the predictions of the BCS model in the usual statistics.

\section{Comparison with Experimental Results}
\label{Exp}

In general, usual superconductors exhibit exponential
behavior in the low temperature regime for several
macroscopic measurable functions. As we will see below, the BCS model
with its constant energy gap for a $ s $-wave
superconductor ($ l =  0 $) display such behavior
in the standard statistics.
On the other hand, the so-called exotic superconductors present power law
behavior and it is believed that different pairing states may be
responsible for such behavior. One of the most interesting results of
the present paper is that, even for $ s $-wave pairing state, it is
possible to find power law behavior for different functions in the
$ T \ll T_{c} $ regime once $ q \not= 1 $.

In this section we compare experimental results for specific heat,
ultrasonic attenuation and tunneling experince for some usual superconductors with
different theoretical predictions for $ q =1 $. We show that small
deviations with respect to $ q = 1 $ provide better agreement when
confronted to data, even regarding the approximations involved in the
calculations. Furthermore, we expect that the
same $ q $ may fit the avaiable data for any different experiments of a given
material.

\subsection{Specific Heat}
\label{Cv}

In an unitary volume, the eletronic specific heat in the
superconductor state is
\begin{equation} \label{Ce}
C^{s}_e =
2 k_B \beta^2
\sum_k
\left( E_k^2 + \frac{\beta}{2}  \frac{d}{ d\beta } \Delta^2 \right)
\left[ -\frac{d}{d(\beta E_k)} f_q(\beta E_k) \right] \, ,
\end {equation}
which may be interpreted as the specific heat associated to 
the quasi-particles with energy $ E_{ k } $, added to the contribution 
of the temperature dependent gap. Notice that when 
$ q \rightarrow 1 $, 
$ -\frac{d}{dx}f_1(x) =  f_1(x)\left( 1 - f_1(x) \right) $
and one recover the usual expression for a superconductor in the BG statistics.

The ``normalized'' jump
$ (\, C^{s}_{e} - C^{n}_{e} \,) / C^{n}_{e} $ may be measured and 
compared with theoretical predictions. The BCS theory with standard statistics
predicts the result of 1.43 for the specific
heat jump, and usual superconductors exhibit some deviations
from this value as it is shown on Table \ref{Cvtable} for several materials.
We calculated numerically
$ (\, C^{s}_{e} - C^{n}_{e} \,) / C^{n}_{e} \left)_{ T_c } \right. $
as a function of $ q $ and therefore we can estimate the pertinent value
of $ q $ for the superconductors listed. Notice that $ C^{n}_{e} $ was considered
as a constant parameter in order to give the value of 1.43
at $ q = 1 $ for the specific heat jump.
Notice, from Table \ref{Cvtable}, that the appropriate $ q $-value for Sn is 0.995.
we shall see below that numerical calculations for Sn using $  q \sim 0.99 $
show better agreement with measurements for the specific heat in the low temperature regime,
ultrasonic attenuation rate and tunneling experiences.
The fact that the same $ q $ is able to give, at least qualitatively, a better
agreement with different measurements serves as motivation to further
application of nonextensive statistics in order to explain the anomalous features
presented in some superconductors.
{\centering
\begin{table} \label{Cvtable}
\begin{center}
\begin{tabular}[width=0.9\columnwidth]{ |c||c |c| }
\hline
       &        &      \\ 
       &   $\frac{ C^{(s)}_e - C^{(n)}_e }{ C^{(n)}_e } \left|_{T_c}
\right. $ &  $q$ \\
       &        &       \\  
\hline
\hline
         &                   &      \\ 
Al       &   1.29 - 1.59     &  1.03 - 0.996\\  
Zn       &   1.3             &  1.03        \\
In       &   1.73            &  0.993    \\
Mo       &   1.28            &  1.03     \\
Nb       &   1.87            &  0.975    \\
Sn       &   1.6             &  0.995    \\ 
Ta       &   1.59            &  0.995    \\
Tl       &   1.5             &  0.997    \\  
         &                   &           \\ 
\hline 
\end{tabular} 
\end{center}
\caption{Display of the values $q$ found when numerical plot is compared
to the specific heat jump measured experimentally\cite{Parks}.
$ C^{(s)}_e $ was calculated numerically in the weak-coupling
limit. Notice that these usual superconductors are well described by $q \approx 1 $. }
\end{table} }

Thus, according to the above discussion, we show the specific heat for Sn and V
in Fig. \ref{CvLTSn}. From Table \ref{Cvtable}, we expected that $ q = 0.995 $
should reproduce better data for Sn than $ q = 1 $.
We see also that the specific heat measurements for vanadium and tin
\cite{Biondietal} are well fitted by the exponential function
$  9.17 e^{-1.5 (\, T_c / T \,) } $ in the temperature region
$ 0.25 < T / T_c < 0.7 $, which is quite remarkable if compared to the prediction
of the usual BCS model, $ 8.5 e^{-1.44 (\, T_c / T \,) } $,
for $ 2.5 < T_c / T < 6 $. Theoretical plots for $ q = 1 $ and 0.99
agree well to the experimental fitting above.
However, one can see in Fig. 1, specially in the insert,
that the agreement for V with $ q = 0.99 $ is better not only at $ T \sim T_c $,
but specially for lower temperatures. At such low temperatures there is an
increasing departure from the standard statistics.

By the same token, on Fig. 2 we show several data for the specific heat for Al and Zn.
They both deviate from the usual BCS calculations in the low temperature regime. The data of
Phillips\cite{Phillips} for
Zn is better fitted by a $ T^3 $-law in the low temperature regime. According to our estimations,
shown in Table \ref{Cvtable},
the Al and Zn data should be fitted by $ q = 1.03 $.
Indeed, one can see in Table \ref{Cvtable}
that the calculations with $ q = 1.03 $ agree much better with the experimental
results of several groups than the usual statistics. We can also see in Fig. 2
the tendency of the specific heat to deviate from a exponent to a power law behavior
even with such small value of $ q = 1.03 $. This behavior is even stronger with a much larger
value of $ q $, like $ q = 1.2 $ for instance, although such calculations will be
presented elsewhere.

\subsection{Ultrasonic Attenuation Rate}
\label{alphasec}

In this section we treat the problem of acoustic waves perturbing
the electronic system described by the diagonalized BCS reduced Hamiltonian.
We will consider only longitudinal waves and the temperature interval, the frequency
and wave vector are related by the inequality $ w < v_F p < \Delta / \hbar $.
As usual, we assume that the matrix element of the perturbation depends on
$ {\bf k}' - {\bf k} $. In a typical ultrasound experience the frequency is
less than $ 10^9 $ Hz, hence $ \hbar w \ll k_B T $, and in the limit of 
very low frequencies one get that the normalized ultrasonic attenuation
rate within Tsallis' statistics is
\begin{equation} \label{alpha}
\frac{ \alpha_s }{ \alpha_n}  =
- \int_{ -\infty }^{ \infty } \frac{ \partial f_q }{ \partial E}
= 2 f_q(\Delta ) \, .
\end{equation}
Notice that $ \alpha_ s / \alpha_n $ also has exponential decrease when
$ q \rightarrow 1 $ in the low temperature regime. Such behavior is experimentally
verified in several usual superconductors, although plots of experimental data tend to
fall off somewhat more rapidly just below $ T_c $. Data of Morse and Bohm\cite{Morse}for Sn
are a example, Fig.3, and are compared with theoretical plots for
$ q = 0.99 $ and 1. Again, $ q = 0.99 $ is in better agreement with experimental results.
An interpretation for the discrepancy between the usual theory and the data was given in terms
of the gap function, which, compared with it value at $ T = 0 $, rises slightly too rapidly
as the temperature decreases just below $ T_c $. Tunneling measurements for
Sn by Morse and Bohm\cite{Giaever} are consistent with such feature. On Fig. 4
we show their results for Sn and In with the calculation with $ q = 0.99 $ and $ 1 $.
We can easily see that $ q = 0.99 $ is in much better agreement than the usual $ q = 1 $
results. Thus in the present
interpretation this result is attributed to the nonextensivity of the system.

\section{Conclusions}
\label{Conc}

We conclude this paper pointing out our new findings:
\begin{itemize}

\item We have show that the BCS potential widely used in the description of superconductors
is nonextensive. Following this result we have developed a BCS theory with
nonextensive entropy which depends on the index $ q $. We have, therefore,
derived some thermodynamic quantities that can be compared with the experiments.

\item Using the experimental values of the specific heat, we estimated
the nonextensive index $ q $ for several superconductors. Explicit calculations
with such values of $ q $ for several materials are indeed in much better agreement
with the present experimental results than the usual values. In particular for Sn
we have estimated $ q \approx 0.99 $ and have shown that with such value we reproduced
with a better agreement the experimental results of the specific heat, ultrasonic attenuation
and tunneling experiments. Similar calculations were performed for Al and Zn. The fact that three
well-known properties are fitted by the same $ q \not= 1 $ rules out a possible deviation due
to the approximations on the calculations.

\end{itemize}

Furthermore, we have seen that as $ q $ increases, the exponential behavior of the specific
heat changes in a power law behavior. This is a novel result and it means that
even if there is a constant gap in the neighborhood of a Fermi surface,
we can get a power law behavior once we use Tsallis statistics with $ q \not= 1 $.
Several new superconductors, like high temperature superconductors for instance,
exhibit such power law behavior
which is attributed to a non-constant d-wave order parameter. At the moment we are developing a
non weak-coupled theory of superconductivity with Tsallis statistics to deal with these and others
exotic superconductors with thermodynamical quantities which exhibit a power law behavior.

\centering{ \sc Acknowledgments}

We thank Conselho Nacional de Desenvolvimento Cient\'{\i}fico e Tecnol\'ogico-CNPq-Brasil e
FAPERJ for partial financial support.

\newpage

{\centering
\begin{figure} \label{CvLTSn}
\begin{center}
\end{center}
\caption{ Comparison between the fitting of the low temperature
electronic specific heat of Tin,
$ C_e^{ (s) } = 9.17 exp[\,-1.5 (\, T / T_c \,)\,]\;[\, \gamma T_c \,] $,
and the theoretical plots for
$ q = 1 $ and $ 0.99 $. Both values of $ q $ describe very well the
experimental results, but for V in particular, regarding the interval of temperature
$ 0.25 < T / T_c < 0.7 $, a better agreement was found with $ q = 0.99 $. }
\end{figure} }

{\centering
\begin{figure} \label{CvAl}
\begin{center}
\end{center}
\caption{ Experimental data for the specific heat of Al and
Zn \cite{Boorse}. Notice that the measurements by Phillips\cite{Phillips}, Goodman\cite{Goodman}
e Zavaritskii\cite{Zavaritskii}
for the Aluminum are well described by the generalized BCS model
with $ q = 1.03 $, the same value furnished in Table \ref{Cvtable}.
Also notice that data obtained by Phillips\cite{Phillips} for Zinc
are very well described by $3 ( T_c/T)^3 $ in the low temperature
regime.  }
\end{figure} }

\begin{figure}
\label{alphaSn}
\caption{Ultrasonic attenuation rate calculated for different values of $ q $
and compared to experimental data for Sn\cite{Morse}. A better agreement
was found with $ q = 0.99 $, as proposed by measurements of the specific heat jump.}
\end{figure}

\begin{figure}
\label{Gapexp}
\caption{Comparison between the gap
energy from tunneling measurements for Tin and Indium\cite{Giaever}
and theoretical plots using $ q = 1 $ and $ 0.99 $. Notice that the gap function related to
 $ q = 0.99 $ have better qualitative agreement with experimental results.}
\end{figure}

\end{document}